# Non-linear effects and Joule heating in I-V curves in manganites

Silvana Mercone, Raymond Frésard, Vincent Caignaert, Christine Martin, Damien Saurel, Charles Simon,

Laboratoire CRISMAT, UMR 6508 ENSICAEN et CNRS, 14050 Caen, France.

Gilles André,

Laboratoire Léon Brillouin,

Centre Etudes de Saclay, CEA-CNRS, 91191 Gif/Yvette, France

Philippe Monod,

Laboratoire de physique du solide, ESPCI, 75005 Paris, France.

François Fauth,

ESRF, 38042 Grenoble Cedex, France.

**Abstract :** We study the influence of the Joule effect on the non-linear behavior of the transport I-V curves in polycrystalline samples of the manganite $Pr_{0.8}Ca_{0.2}MnO_3$ by using the crystalline unit cell parameters as an internal thermometer in X-ray and neutron diffraction. We develop a simple analytical model to estimate the temperature profile in the samples. Under the actual experimental conditions we show that the internal temperature gradient or the difference between the temperature of the sample and that of the thermal bath are at the origin of the non-linearity observed in the I-V curves. Consequences on other compounds with colossal magnetoresistance are also discussed.

**Introduction:**

The mixed valence manganese based perovskites are the object of an intense activity [1] due to the existence of electronic nanophase separation and related colossal magnetoresistive properties. Among the fascinating properties, non linear electric response in I-V curves is a very intense subject of recent discussions [2-12]: among the different compounds, the family $Pr_{1-x}Ca_xMnO_3$ was extensively studied [10] and different interpretations were proposed including melting of the charge ordering [8], modification of the percentage or of the shape of the phase separation by the current, amplified by the percolation phenomenon [13,14] or the depinning of possible charge density waves related to the charge ordering [9,10]. It is beyong the scope of this paper to discuss the details of these interpretations, but one of the most puzzling cases is $Pr_{0.8}Ca_{0.2}MnO_3$ at 100K, in which no charge ordering or phase separation was observed in the structure [15] though a strong non-linear effect in the resistivity was found [5], together with a sharp drop of the magnetization.

In reference [5], as in most of the papers published on these topics, a special care was taken in order to rule out the possibility of trivial Joule effect, which can be a possible interpretation of these data. Estimations of the Joule effect in bulk samples were proposed in the publications. However, we will show in the present paper that, in most of the experimental situations, these estimations were not correct and the dominant effect is indeed Joule heating. The reasons of this unusual mistake are that the thermal conductivity of these compounds is quite low due to the presence of cationic disorder on the (Pr,Ca) site which prevents easy phonon propagation (about 1 $Wm^{-1}K^{-1}$ at 100K [16]) and that the dissipated electric power is usually quite high in these measurements (about $10W/cm^3$ ). Under these conditions, the internal gradient can dominate the external temperature difference between the sample and the thermal bath. Then, the control of the surface temperature is not a good test. Moreover, in most of the thermal estimations, the current is assumed to flow homogeneously, and we will show that is not true in such badly thermally conducting samples. The relevance of this type of instabilities has been very well known for a long time in other systems[17]. The situation of the bulk samples is very different from that of the thin films in which most of the temperature gradient is in the substrate [18, 19].

In this paper, we present in the first part the analytic calculation of the internal gradient of temperature and current density in two different situations –flat sample and circular sections. To prove the validity of our estimations, it was necessary to measure the temperature inside the sample, in addition to the surface temperature. In order to do that, we have used an internal thermometer, i.e., the structural cell parameters measured during diffraction measurements performed at the neutron Léon Brillouin and the X-ray ESRF facilities. Moreover, neutron scattering allows simultaneous measurement of the increase or decrease of the magnetization of the sample.

**Calculation of the Joule effect**

The determination of both the Joule effect and current distribution profile for a given sample is known to be cumbersome, in particular for a semiconductor [20]. In contrast, we have found that the analytic calculation can be carried out completely under different conditions. For temperatures between 100K and 200K the thermal conductivity is only weakly temperature dependent, taking the value $K_{th}$=1 $Wm^{-1}K^{-1}$ [15], while as shown in Fig.1 the temperature dependence of the resistivity of the manganite sample can be fitted by:

$R=R_0 \exp -\{\beta\ (T/T_0-1)\}$ (1)

At first glance the overall behavior of the resistivity of the investigated material in the temperature range (4-300K) is semiconductor-like. However a more careful analysis reveals that the temperature dependence of the resistivity is best represented by Eq. 1 in the range (100-200K).

First we proceeded with a model where:

(i) The sample of length L and diameter t is threaded by a current flowing in the x direction. A voltage V applied in this direction creates an electric field E which is constant everywhere (see Fig. 2).

(ii) The external surface at r=t/2 is in perfect thermal contact with the thermal bath at temperature $T_0$. There is no singularity in the temperature profile at the center r=0 (dT/dr=0).

This condition (ii) is an approximation, which is only valid if the temperature gradient at the interface can be neglected compared to the internal temperature gradient. We have also neglected the effect of the two ends of the sample along the x direction. These are rather good approximations if the sample is long.

Since the electric field is homogeneous, we are left with solving the thermal equation [20]

$$d^2T(r)/dr^2+1/r\ dT(r)/dr = - E^2 /\rho(r)K_{th} \qquad (2)$$

together with the boundary conditions $T(t/2) = T_0$ and $dT/dr(r=0)=0$. (Please note that this is the same temperature as the one chosen in Eq. (1) in which it is clearly arbitrary). Using Eq. (1) and introducing the dimensionless quantities $\theta= \beta(T(\zeta)/T_0 -1)$ , $\zeta =r/t$ and

$$C^2 =( tE)^2 \beta/(\rho_0 T_0 K_{th}), \text{ C being the dimensionless electric field} \qquad (3)$$

the thermal equation reads:

$$d^2 \theta (\zeta)/d\zeta^2+1/\zeta\, d\theta (\zeta)/d\zeta = - C^2 \exp \{ \theta (\zeta) \} \qquad (4)$$

Eq. (4) has a unique even regular solution, despite its non-linear character. It reads:

$$\theta (\zeta) = \ln \{32X/(C^2+4X\zeta^2 )^2\} \qquad (5)$$

where $X=(16-C^2 + 4(16-2C^2)^{1/2})$ or $X=(16-C^2 - 4(16-2C^2)^{1/2})$ and $\theta(\zeta)$ is the dimensionless temperature profile for different electric field values C (Fig. 2a). For small values of C and X (not presented in Fig. 2), the temperature gradient is small and smooth. For larger values of X (Fig. 2a) the internal temperature gradient can be very important. Then the current density $j(\zeta) = (E/\rho_0) \exp( \theta (\zeta) )$ can be also very inhomogeneous (Fig. 2b). We can now determine the total current threading the sample $I=t^2 \int 2\pi\zeta\, d\zeta\, j(\zeta)$ to obtain

$$R/R_0 = (C^2/128)\{(X+4C^2)/X)\} \qquad (6)$$

where $I_{cr}$ is the value of I which corresponds to the maximum of the corresponding electric field:

$$I_{cr} = (8\pi t/5)(2K_{th}T_0/\beta\rho_0)^{1/2}. \quad (7)$$

While R(I) is a monotonic function of I, this does not hold for the voltage V(I). On the contrary, it is limited by a critical value which corresponds to $C=2\sqrt{2}$ above which the thermal equation has no solution. It corresponds to the maximum voltage of the I-V curves. For smaller values of V, there are two solutions, each corresponding to a different value of the current I.

In this second part, we have also solved a model where:

(iii) The sample of length L and width w is threaded by a current flowing in the x direction. A voltage V applied in this direction creates an electric field E which is constant everywhere (Fig. 2).

(iv) The external surfaces at z= t/2 or z=-t/2 are in perfect thermal contact with the thermal bath at temperature $T_0$.

Since the electric field is homogeneous, we are left with solving the thermal equation [20]

$$d^2T(z)/dz^2 + dT(z)/dz = - E^2 /\rho(z)K_{th} \quad (8)$$

together with the boundary conditions $T(t/2) = T(-t/2) = T_0$. Using Eq. (8) and introducing the same dimensionless quantities the thermal equation is given by:

$$d^2 \theta(\zeta)/d\zeta^2 + d\theta(\zeta)/d\zeta = - C^2 \exp\{ \theta(\zeta) \} \quad (9)$$

Eq. (9) also has a unique even regular solution which is:

$$\theta(\zeta) = -2 \ln\{\cosh(X\zeta/2)/\cosh(X/4)\}$$

where X is obtained from $C=(X/\sqrt{2}) / \cosh(X/4)$ (10)

In Fig. 2, we have superimposed these results compared to those of the first geometry. The results are only slightly different. We can now determine the total current threading the sample $I=V/R_0 \int d\zeta\, e^{\theta(\zeta)}$ to obtain

$$i = I/I_{cr} = \sinh(X/4) \qquad \text{with } I_{cr} = 2\,w\,(2K_{th}T_0/\beta\rho_0)^{1/2}. \qquad (11)$$

And then

$$R/R_0 = \ln\{ i + (1+i^2)^{1/2} \} / i\,\{(1+i^2)^{1/2}\} \qquad (12)$$

In Fig. 3, we have superimposed the result of the two calculations for $R/R_0(i)$. For the first model, the data are calculated from the two implicit equations I(C)=C/R(C) and R(C) (Eq. 6). In these reduced units, there is no adjustable parameter. This non-linear calculation, which is very similar to what is usually observed in previous publications [2-12], is only due to Joule effect usually neglected up to now. In order to understand the physical origin of the discrepancy between this estimation and the ones previously published, let us look at the plot of the temperature and the current profile across the sample (Fig. 2). One can see that, due to the non-linear temperature dependence of the resistivity, the current flows mainly in the center of the sample, which is at a higher temperature. This effect which was not taken into account in previous estimations, is at the origin of this discrepancy.

From these calculations, the non linear shape observed in the experimental data can be due to this Joule effect. The shape of the sample is not the relevant parameter for the calculation of $I_{cr}$ ($(8\pi t/5)$ is replaced by 2w), but is very important to calculate in detail the temperature profile and the shape of the R(I) curve. The important point is now to quantify the parameter $I_{cr}$ of the model.

**Experimental aspects:**

In order to quantify the parameter $I_{cr}$, we have cut two samples from the same batch of $Pr_{0.8}Ca_{0.2}MnO_3$ composition. The non-linear electric response was very similar to what was found in single crystalline samples of the same composition previously published [10]. The first sample (a rod of 8mm diameter and 25mm long) was measured by the G4.1 diffractometer of the Laboratoire Léon Brillouin. The sample is held in a classical orange cryostat with a helium pressure of 50 mbar in which the neutron beam covers roughly the whole sample. The second sample was a flat plate of 5x10 $mm^2$ and 0.5 mm thick. Here the X- ray beam also covers the whole sample and crosses the sample. The heat transfer was done by helium gas at atmospheric pressure.

The diffraction patterns have been recorded as functions of temperature by first cooling the sample down to 100K, and then increasing the temperature. The sample was then cooled down to 100K, and the current was increased. The diffraction patterns were recorded after reaching equilibrium in the resistance measurement.

Under similar experimental conditions, a measurement was performed with a small external thermometer (thermocouple) attached onto the surface in order to determine the gradient between the surface of the sample and the thermal bath. This experiment was performed at 100K in the experimental chamber of the Quantum Design SQUID magnetometer MPMS5 on a sample with 0.6 $cm^2$ surface area. It was shown that the temperature at the surface of the sample increases by about 10K when a power of 0.02W is dissipated at 100K. Since the surface is about 0.6 $cm^2$, the heat transfer is about 30 W $K^{-1}$ $m^{-2}$ to the thermal bath.

## Experimental results

The temperature dependence of the resistivity was measured on both samples between 100K and 200K. The results are presented in Fig. 1. The resistance can be fitted by Eq. 1 with $\beta$=30, $T_0$=100 K, $\rho_0$ = 128 Ωm. The non-linear electric response was also measured at 100K and found to be very similar to what was previously published [10] (Fig. 4). The critical current was found to be 0.2 (10) mA for the slab (rod) sample. It should be pointed out that the results of the calculations are obtained assuming that the surface remains at the temperature of the bath, $T_0$. The difference between the temperature on the surface and $T_0$ can be estimated by calculating the power dissipated at $I_{cr}$ and using the thermal coefficient experimentally measured (30 W $K^{-1}$ $m^{-2}$). The temperature difference is about 10K at $I_{cr}$ in both samples, which must be taken into account in the calculations. When determining the corresponding values from the model of the previous section with the geometrical factors, one gets $I_{cr}$ = 0.2(16) mA for the slab (rod) sample, respectively, in good agreement with experimental values.

The neutron scattering diffraction patterns were fitted using the Pnma structure. From these results, we present the volume of the unit cell of the Pnma structure, v, in Fig. 5a. Even though the variation of the volume is very small in the temperature range (80-150K) one clearly sees that the thermal expansion abruptly jumps slightly below the Curie temperature. Here the temperature dependence of both v and the magnetic moment, M, are measured on a zero field cooled sample under an increase of temperature. The current dependence of both the volume of the unit cell and the one of the magnetic moment shown in Fig. 5b when the experiment is repeated with an electric current flowing through the sample and cooled in a bath at temperature $T_0$. At the smallest current reported here (1 mA), both v and M coincide with their value measured at 100 K in the absence of a current. Knowing both v(T) and v(I) allows for obtaining T(I). The resulting data, for the slab sample, is compared to our model on Fig. 6. Using $I_{cr}$ = 0.2 mA from above, we see that the data and the model are in excellent agreement. The latter can only be obtained taking Joule effect into account: indeed the resulting temperature increase can be quite substantial, reaching 45 K at I = 10 mA, under which circumstances the magnetic moment naturally vanishes.

Since there may be a sizeable temperature gradient in the sample, an observable broadening of the lines in the diffraction patterns may be expected, even though the lattice parameters vary by about 2% between 100 K and 150 K. In these experiments, resolution was quite poor, in particular due to the poor crystallinity of the samples, and no broadening of the peaks was observed.

This conclusion has other consequences: the existence of two solutions for a given electric field below a critical value can be at the origin of instabilities observed when the measurement is performed at a constant voltage [7]. When the current is fixed, there is a single solution, but highly non-linear in voltage, with a strong current density gradient across the sample. This effect can also explain why the magnetization decreases strongly (when the the temperature of the center part of the sample crosses that of the magnetic transition).

We have observed the same effect in a charge ordered sample of composition $Nd_{0.7}Ca_{0.3}MnO_3$. One should also note that in this case, the magnetization increases in this temperature range as it was reported previously to be an indication of the increase of the ferromagnetic fraction inducing the percolation [9]. This is only due to the fact that the magnetization increases with temperature in this temperature range.

In the more complex case of the colossal magnetoresistive samples in which there is a maximum in the curve resistivity versus temperature, it should also be pointed out that some instabilities may occur, jumping from the unstable situation - in which dissipation increases as the temperature increases - to a stable situation in the other part of the resistivity curve. We have published this discussion in a previous paper on thin films compounds [19].

This solution in which the internal gradient is dominant compared to the coupling with the thermal bath is valid for quite a wide range of sample thicknesses. Depending on the thermal coupling to the thermal bath, there exists a critical thickness, which limits the validity of this model. For a thin film, for example, one should compare the internal thermal gradient to the gradient across the substrate. Since the thermal conductivity of a typical oxide substrate of thickness 0.5 mm is 0.15 $WK^{-1}cm^{-1}$ , the critical thickness is about 10 μm. Below this thickness, classical models used to calculate the film temperature using the gradient across the substrate are correct as long as the width of the sample remains large compared to its thickness [18, 19]. In the case of microbridges (w not larger than 10μm), the edge effects not usually taken into account should of course also play an important role.

It should be also pointed out that the Joule effect explains why, despite the negative differential resistance observed here, the measurements performed by superimposing a small AC current on the main DC current always provides positive AC resistance since the temperature is roughly constant in this configuration.

**Concluding remarks**

In conclusion, we have shown that the non linear behavior of the electric response of the $Pr_{0.8}Ca_{0.2}MnO_3$ composition is due to Joule heating. A combination of low thermal conductivity and high non linear behavior of the temperature dependence of the resistivity is at the origin of this effect. Since the number of published papers on this subject is quite large, we did not review all of them. However, we want to point out as a conclusion that most of the arguments to estimate the Joule heating in the previous publications (including ours) have to be reinvestigated.

Acknowledgments: We acknowledge interesting discussions with M. Weissman, A. Wahl, W. Prellier, V. Hardy and P. Monceau and H. Eng for helping us in writing the manuscript. SM acknowledges support from European Community and DS for the région Basse Normandie.

Figure Captions:

Fig. 1: Temperature dependence of the resistance of the rod $Pr_{0.8}Ca_{0.2}MnO_3$ polycrystalline sample.

Fig. 2: Spatial repartition of the temperature and current density normalized to its value on the surface for different values of reduced electric field C as calculated in the two models (rod: dashed lines, slab: plain lines). One should notice the strong increase of the current density at the center of the sample in the non linear regime. In the inset, the two sample shapes (rod and slab) are presented.

Fig. 3: Calculated current dependence of the sample resistance in the case of a rod (plain line) and slab (dashed line) samples.

Fig. 4: Resistance versus current obtained in the diffraction measurement at 100K for the rod $Pr_{0.8}Ca_{0.2}MnO_3$ sample.

Fig. 5: a) Temperature dependence of the volume of the unit cell (squares), and of the magnetic moment (circles), for the rod sample. The data are recorded under an increase of temperature. The lines are guides for the eyes, only. b) Current dependence of the volume of the unit cell (squares), and of the magnetic moment (circles), for the rod sample. The data are recorded under a decrease of the current. The lines are guides for the eyes, only

Fig. 6: Comparison between temperatures: calculated (full line) and measured from diffraction experiment (crosses) as a function of normalized applied current.

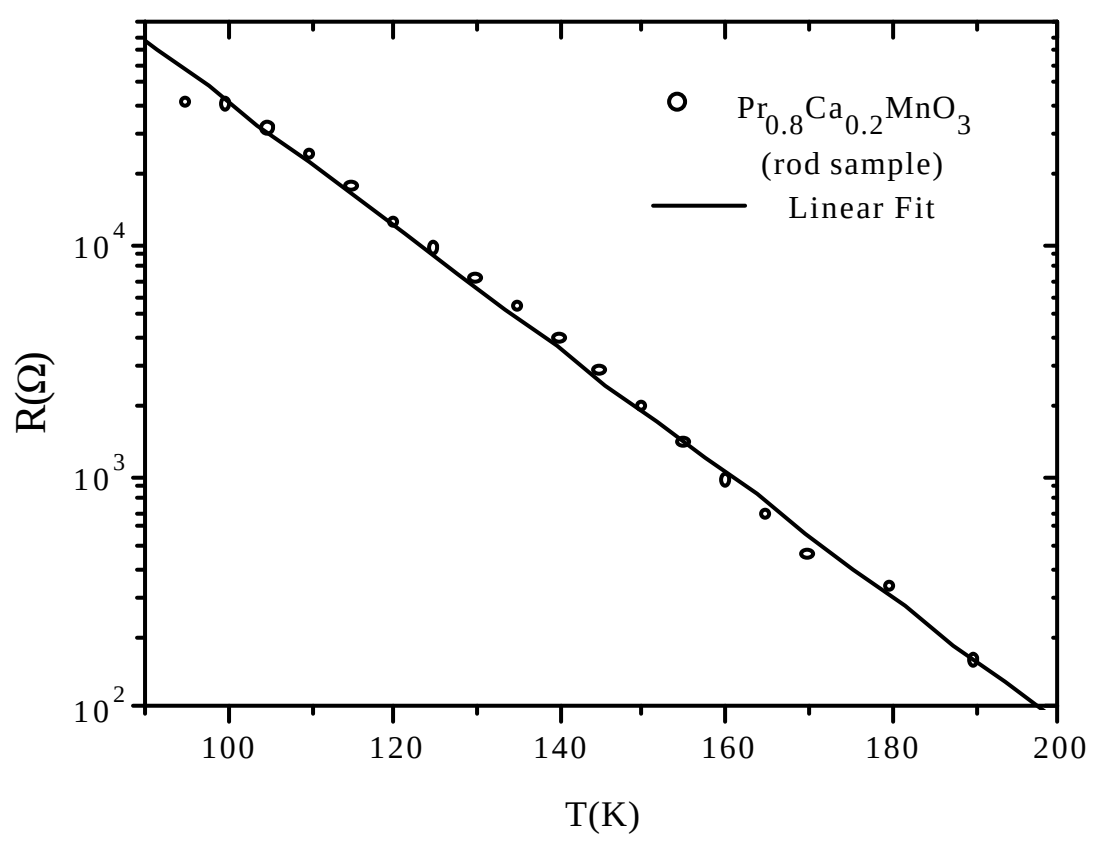

S.Mercone *et al.*

Figure 1

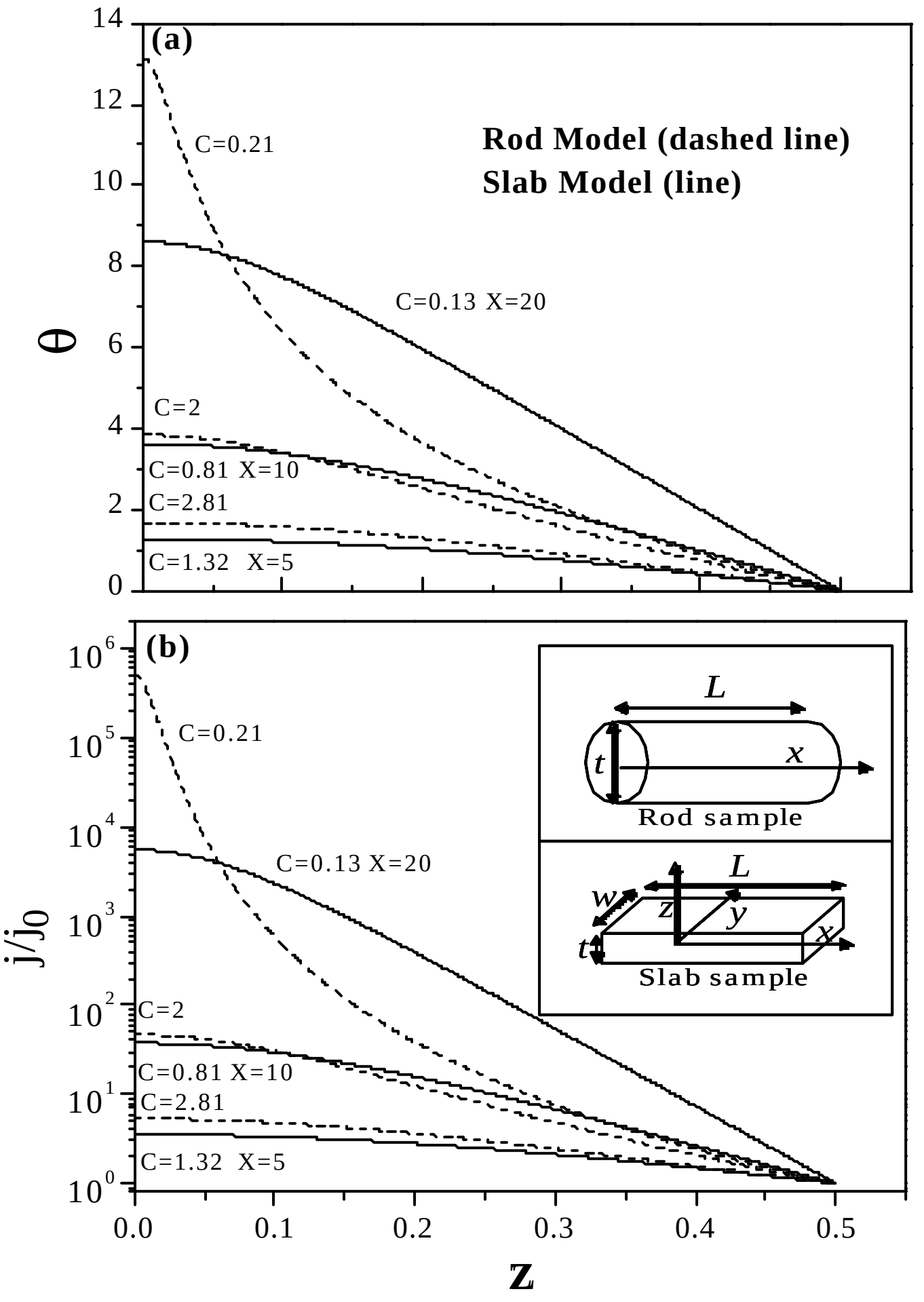

S.Mercone *et al.*
Figure 2

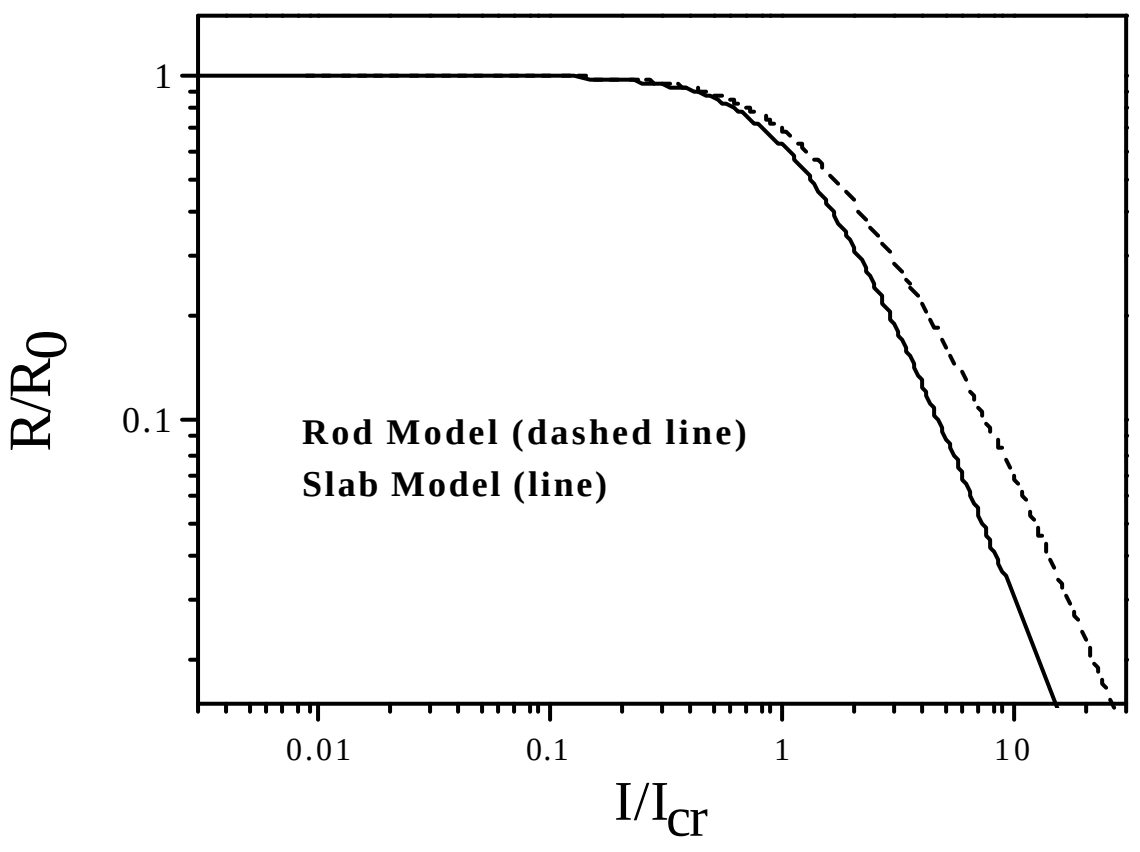

S.Mercone *et al.*

Figure 3

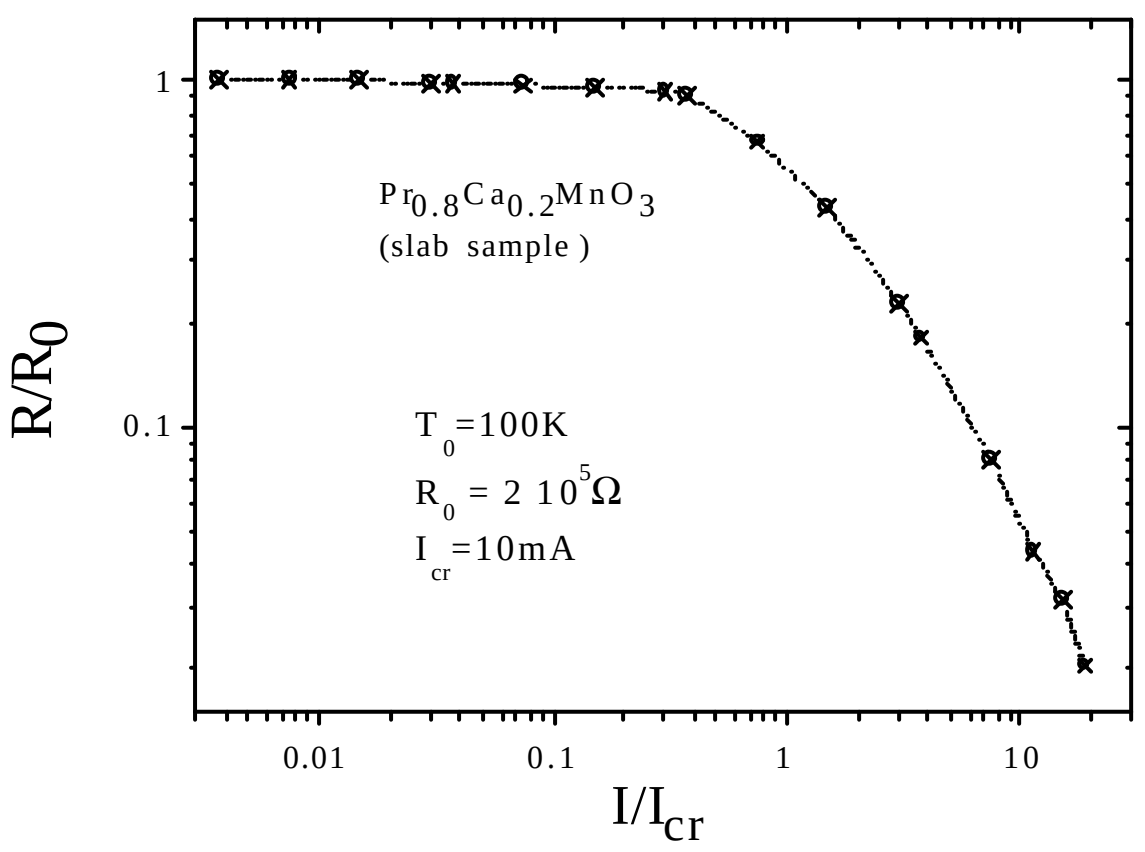

S.Mercone *et al.*

Figure 4

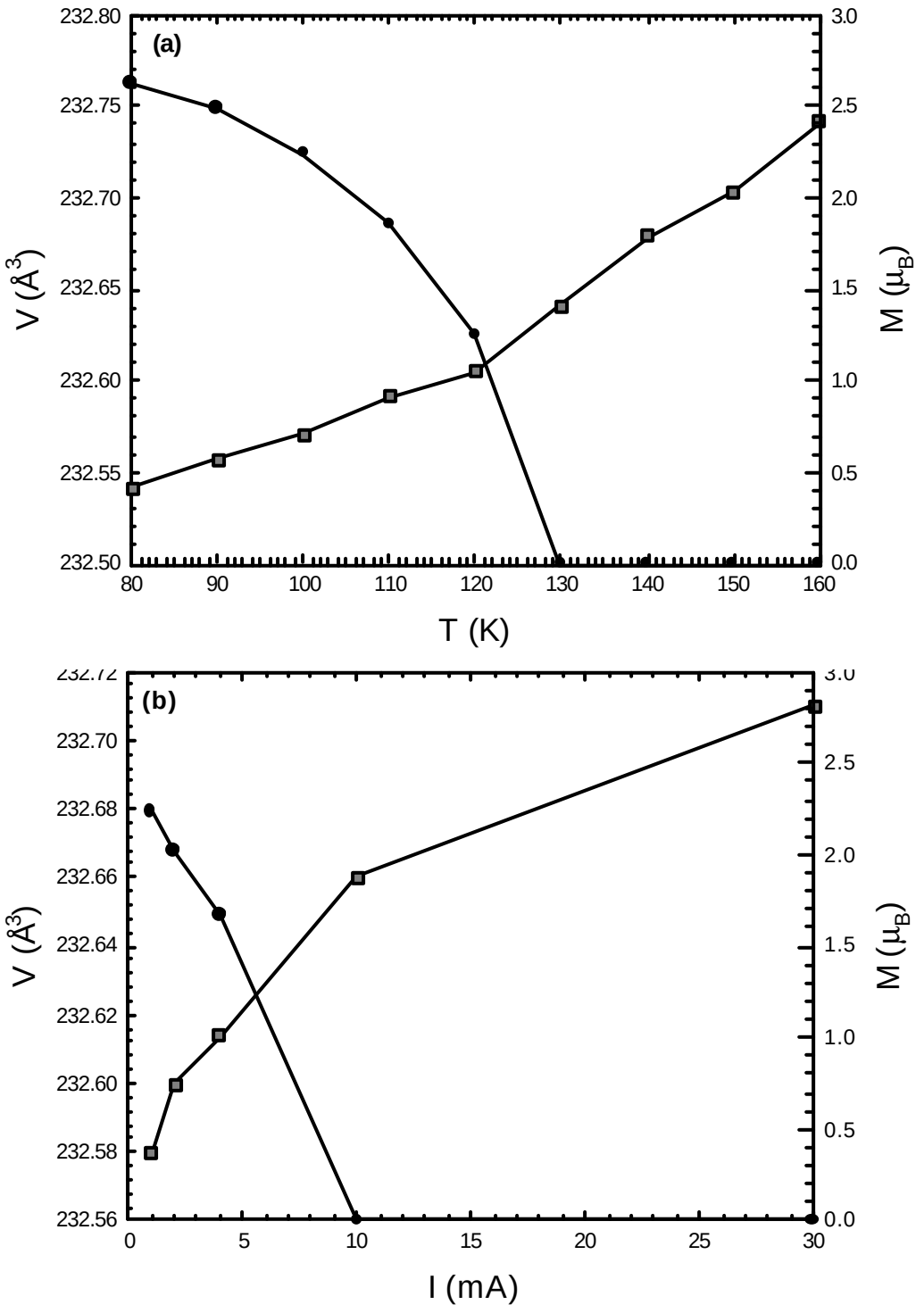

S. Mercone *et al*

Figure 5

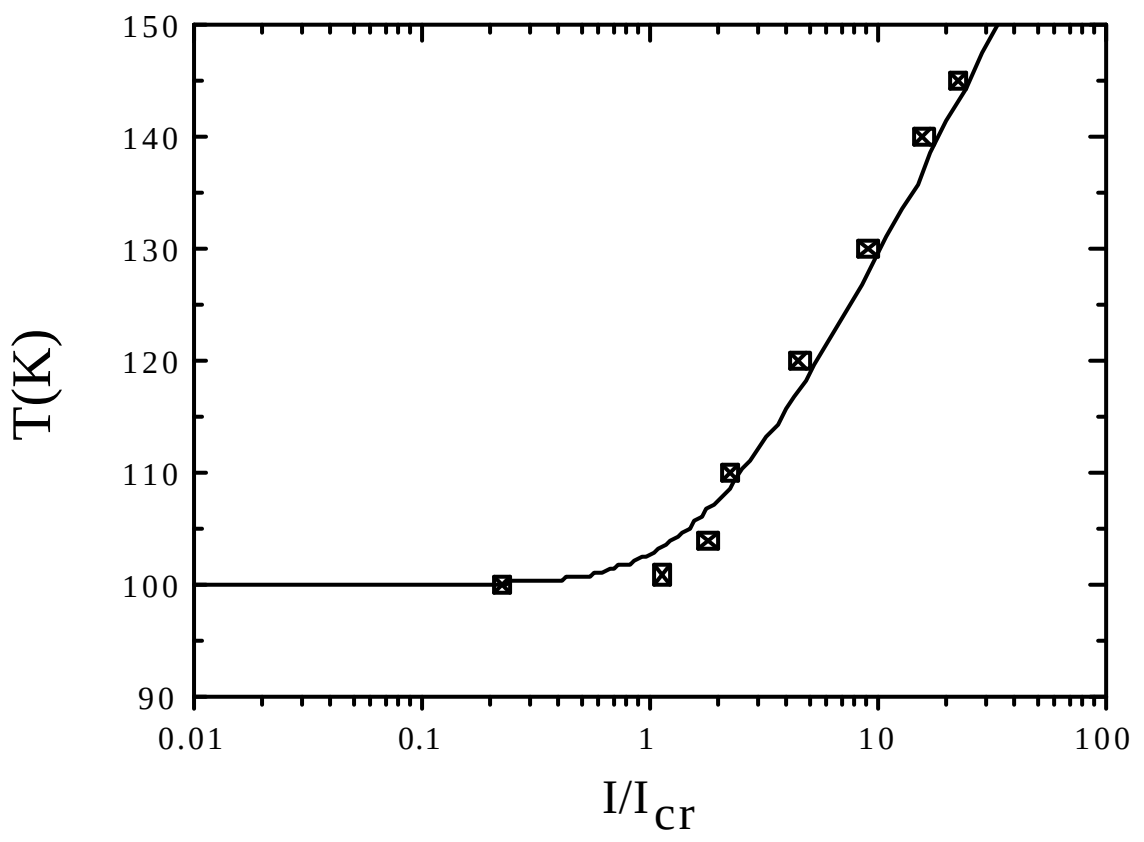

S.Mercone *et al.*
Figure 6